\documentclass[a4paper]{article}

\usepackage[letterpaper]{geometry}
\usepackage[parfill]{parskip}
\PassOptionsToPackage{numbers, compress}{natbib}
\usepackage[numbers, compress]{natbib}
\usepackage{xr-hyper,refcount}

\usepackage{graphicx}
\usepackage[utf8]{inputenc} 
\usepackage[T1]{fontenc}    
\usepackage[hyphens]{url}  
\usepackage{tabularx}
\usepackage{booktabs}       
\usepackage{amsfonts}       
\usepackage{nicefrac}       
\usepackage{microtype}      
\usepackage[parfill]{parskip}
\usepackage{algorithm,algorithmic}
\usepackage{amsmath,amsthm,amssymb,bbm}
\usepackage{mathtools}
\usepackage{cases}
\usepackage{comment}
\usepackage{subcaption}
\usepackage{algorithm,algorithmic}
\usepackage{color}
\usepackage{appendix}
\usepackage{xspace}
\usepackage{enumitem}
\usepackage[english]{babel}
\usepackage{authblk}

\usepackage[colorlinks,citecolor=blue,urlcolor=blue,linkcolor=blue,linktocpage=true,breaklinks]{hyperref}
\hypersetup{breaklinks=true}

\usepackage[hyphenbreaks]{breakurl}
\usepackage{cleveref}
\crefformat{equation}{(#2#1#3)}
\crefrangeformat{equation}{(#3#1#4) to~(#5#2#6)}
\crefname{equation}{}{}
\Crefname{equation}{}{}

\usepackage[hyperpageref]{backref}
\renewcommand*\backref[1]{\ifx#1\relax \else (Cited on #1) \fi}

\crefname{definition}{\textbf{definition}}{definitions}
\Crefname{definition}{Definition}{Definitions}
\crefname{assumption}{\textbf{assumption}}{assumptions}
\Crefname{assumption}{Assumption}{Assumptions}
\definecolor{maroon}{RGB}{192,80,77}

	\AtEndDocument{\refstepcounter{theorem}\label{finalthm}}

\usepackage{mdwlist}

\usepackage[utf8]{inputenc} 
\usepackage[T1]{fontenc}    
\usepackage{url}            
\usepackage{booktabs}       
\usepackage{amsfonts}       
\usepackage{nicefrac}       
\usepackage{microtype}      
\usepackage{tikz}
\usepackage{tabularx}
\usetikzlibrary{automata,decorations.markings,positioning,arrows}

\usepackage{natbib}
\usepackage{amsmath}
\usepackage{bbm}

\usepackage[normalem]{ulem}

\usepackage{mathtools}

\usepackage{amssymb,amsmath,amsthm}
\usepackage{soul}
\usepackage{url}            
\usepackage{xcolor}
\usepackage{refcount}

\usepackage{floatpag}
\usepackage{float}

\usepackage{multirow}

\newcommand{\xhdr}[1]{\vspace{1mm} \noindent\textbf{#1.}}
\newcommand{\qt}[1]{\textit{``#1''}}

\usepackage[]{mdframed}

\begin{document}

\title{Red-Teaming for Generative AI:\\Silver Bullet or Security Theater?}

\author {
    Michael Feffer, Anusha Sinha, Wesley H. Deng, Zachary C. Lipton, Hoda Heidari\\
    Carnegie Mellon University\\
    \href{mailto:mfeffer@andrew.cmu.edu}{\nolinkurl{mfeffer}}@andrew.cmu.edu, \href{mailto:asinha@sei.cmu.edu}{\nolinkurl{asinha}}@sei.cmu.edu,\\
    \{\href{mailto:hanwend@andrew.cmu.edu}{\nolinkurl{hanwend}}, \href{mailto:zlipton@andrew.cmu.edu}{\nolinkurl{zlipton}}, \href{mailto:hheidari@andrew.cmu.edu}{\nolinkurl{hheidari}}\}@andrew.cmu.edu
}

\date{}

\maketitle

\begin{abstract}
  In response to rising concerns surrounding the safety, security, and trustworthiness of Generative AI (GenAI) models, 
practitioners and regulators alike have pointed to \emph{AI red-teaming}
as a key component of their strategies for identifying and mitigating these risks.
However, despite AI red-teaming's central role in policy discussions and corporate messaging,
significant questions remain about what precisely it means, 
what role it can play in regulation,
and how it relates to conventional red-teaming practices
as originally conceived in the field of cybersecurity.
In this work, we identify recent cases of red-teaming activities in the AI industry
and conduct an extensive survey of relevant research literature 
to characterize the scope, structure, and criteria for AI red-teaming practices.
Our analysis reveals that prior methods and practices of AI red-teaming 
diverge along several axes, including the purpose of the activity (which is often vague),
the artifact under evaluation, 
the setting in which the activity is conducted (e.g., actors, resources, and methods), 
and the resulting decisions it informs (e.g., reporting, disclosure, and mitigation). 
In light of our findings, we argue that while red-teaming 
may be a valuable big-tent idea for characterizing GenAI harm mitigations, and that industry may effectively apply red-teaming and other strategies behind closed doors to safeguard AI, gestures towards red-teaming (based on public definitions) as a panacea for every possible risk 
verge on \emph{security theater}.
To move toward a more robust toolbox of evaluations for generative AI, 
we synthesize our recommendations into a question bank 
meant to guide and scaffold future AI red-teaming practices. 

\end{abstract}

\section{Introduction}\label{sec:intro}
In recent years, generative AI technologies,
including large language models (LLMs) 
\cite{Touvron_2023, OpenAI_2023}
image and video generation models 
\cite{Ramesh_Dhariwal_Nichol_Chu_Chen_2022, Rombach_Blattmann_Lorenz_Esser_Ommer_2022,videoworldsimulators2024},
and audio generation models \cite{donahue2023singsong,agostinelli2023musiclm} 
have captured public imagination. 
While many view the proliferation and accessibility of these tools favorably,
envisioning boons to productivity, creativity, and economic growth,
concerns have emerged that the rapid adoption of these powerful models
might unleash new categories of societal harms. 
These concerns have gained credibility owing to several well-publicized problematic incidents
where such AI output text expressing discriminatory sentiment 
towards marginalized groups \cite{mei2023bias, ghosh2023chatgpt, omrani2023evaluating,haim2024s,hofmann2024dialect}, 
created images reflecting harmful stereotypes \cite{Luccioni_Akiki_Mitchell_Jernite_2023,wan2024male},
and enabled the generation of deepfake audio in a fashion that has been likened to \emph{digital blackface} \cite{Feffer_Lipton_Donahue}.
These issues are compounded by the lack of transparency and accountability
surrounding the creation of these models~\cite{birhane2021multimodal,OpenAI_2023,Widder_West_Whittaker_2023}.

In answer to the mounting worry over the safety, 
security, and trustworthiness of generative AI models, 
practitioners and policymakers alike have pointed 
to \emph{red-teaming} as an integral part of their strategies
to identify and address related risks, with the goal of ensuring some notion of alignment with human and societal values \cite{Anthropic_frontier_rt,mrbullwinkle_2023,Bockting_2023}.
Notably, the US presidential Executive Order on the Safe, Secure, and Trustworthy Development and Use of Artificial Intelligence~\citep{House_2023} mentions red-teaming eight times, defining it as follows: 

\vspace{2mm}

\begin{mdframed}
        \qt{The term `AI red-teaming' means a structured testing effort to find flaws and vulnerabilities in an AI system, often in a controlled environment and in collaboration with developers of AI.  Artificial Intelligence red-teaming is most often performed by dedicated `red teams' that adopt adversarial methods to identify flaws and vulnerabilities, such as harmful or discriminatory outputs from an AI system, unforeseen or undesirable system behaviors, limitations, or potential risks associated with the misuse of the system.}
\end{mdframed}
\vspace{2mm}

\noindent The order mandates the Secretary of Commerce and other federal agencies
to develop guidelines, standards, and best practices for AI safety and security. 
These include 
\qt{appropriate procedures and processes, to enable developers of AI, 
especially of dual-use foundation models, to conduct AI red-teaming tests} 
as a mechanism for 
\qt{assessing and managing the safety, security, and trustworthiness of [these] models.}

On one hand, red-teaming appears to call for \emph{the right stuff}: 
find the flaws, find the vulnerabilities, and (help to) eliminate them. 
In this spirit, one might find its inclusion 
in a landmark policy document to be welcome.
On the other, for all of the virtue in its aims, 
red-teaming at this level of description is strikingly vague. 
As noted by \citet{FMF_2023}, 
\qt{there is currently a lack of clarity on how to define `AI red teaming' 
and what approaches are considered part of the expanded role it plays in the AI development life cycle.} 
For example, the definition offered by the presidential executive order 
leaves the following key questions unanswered:
What types of \emph{undesirable behaviors, limitations, and risks} 
can or should be effectively caught and mitigated through red-teaming exercises?
How should the activity be \emph{structured} to maximize 
the likelihood of finding such flaws and vulnerabilities? 
For instance, aside from AI developers, who else should be at the table, 
and what resources should be available to them?
How should the risks identified through red-teaming be \emph{documented, reported,} and \emph{managed}?
Is red-teaming on its own sufficient for assessing and managing 
the safety, security, and trustworthiness of AI? 
If not, what other practices should be part of the broader evaluation toolbox,
and how does red-teaming complement those approaches?
In short, is red-teaming the stuff of policy---the sort of concrete practice
around which we can structure regulatory requirements?;
or is it the stuff of \emph{vibes}---a vague practice 
better suited to rallying than to rule-making?

\xhdr{Methodology} Using publicly available resources, 
we gathered information about recent real-world cases 
of AI red-teaming exercises (Section~\ref{sec:case-study}). 
We emphasize that many of these cases stem from private sector companies who may use other evaluation techniques not shared with the general public. As such, our corresponding analyses and conclusions rest on disclosed details. 
To complement these case studies primarily conducted by industry, we additionally performed an extensive survey 
of existing research literature on red-teaming
and adjacent testing and evaluation methods 
(e.g., penetration testing, jailbreaking, and beyond) 
for generative AI (Section~\ref{sec:survey}).  
We organized the thematic analysis of our case studies and literature survey
around the following key questions:
\begin{itemize}
    \item \textbf{Definition and scope:} What is the working definition of red-teaming? 
    What is the success criterion?
    \item \textbf{Object of evaluation:} What is the model being evaluated? 
    Are its implementation details (e.g., model architecture, training procedure, safety mechanisms) available to the evaluators or to the public? 
    At what stage of its lifecycle 
    (e.g., design, development, or deployment) is the model subjected to red-teaming? 
    \item \textbf{Criteria of evaluation:} What is the threat model 
    (i.e., the risk(s) for which the model is being evaluated)? 
    What are the risks the red-teaming activity potentially missed?
    \item \textbf{Actors and evaluators:} Who are the evaluators? 
    What are the resources available to them 
    (e.g., time, compute, expertise, type of access to model)? 
    \item \textbf{Outcomes and broader impact:} 
    What is the output of the activity? 
    How much of the findings are shared publicly? 
    What are the recommendations and mitigation strategies produced 
    in response to the findings of red-teaming? 
    What other evaluations had been performed on the model aside from red-teaming?
\end{itemize}

To further extend and validate our analysis, we analyzed public comments submitted to a Request For Information (RFI) issued by the National Institute of Standards and Technology (NIST) arm of the Department of Commerce. This RFI sought opinions on points relevant to red-teaming as outlined in the Executive Order.\footnote{See appendix for RFI comment analysis.}

\begin{table}[tbp]
    \thisfloatpagestyle{empty}
    \centering
\caption{Our proposed set of questions to guide future AI red-teaming activities.}
\label{tab:questions}
\begin{tabular}{|l|l|}
\hline
\textbf{Phase}                               & \textbf{Key Questions and Considerations}                                                                                                                                                                                                                                                                                                                                                                                                                                                                                                                 \\ \hline
\multirow{5}{*}{\textbf{0. Pre-activity}}    & \begin{tabular}[c]{@{}l@{}}What is the \textbf{artifact under evaluation} through the proposed red-teaming activity? \\ - What version of the model (including fine-tuning details) is to be evaluated? \\ - What safety and security guardrails are already in place for this artifact?\\ - At what stage of the AI lifecycle will the evaluation be conducted? \\ - If the model has already been released, specify the conditions of release.\end{tabular}                                                                                                      \\ \cline{2-2} 
                                             & \begin{tabular}[c]{@{}l@{}}What is the \textbf{threat model} the red-teaming activity probes?\\ - Is the activity meant to illustrate a handful of possible vulnerabilities? \\\:\:\:\:(e.g., spelling errors in prompt leading to unpredictable model behavior)\\ - Is the activity meant to identify a broad range of potential vulnerabilities? \\\:\:\:\:(e.g., biased behavior)\\ - Is the activity meant to assess the risk of a specific vulnerability? \\\:\:\:\:(e.g., divulging recipe for explosives)\end{tabular}                                                                                                                                                                                                                       \\ \cline{2-2} 
                                             & \begin{tabular}[c]{@{}l@{}}What is the \textbf{specific vulnerability} the red-teaming activity aims to find? \\ - How was this vulnerability identified as the target of this evaluation? \\ - Why was the above vulnerability prioritized over other potential vulnerabilities? \\ - What is the threshold of acceptable risk for finding this vulnerability?\end{tabular}                                                                                                                                                                                       \\ \cline{2-2} 
                                             & \begin{tabular}[c]{@{}l@{}}What are the \textbf{criteria for assessing the success} of the red-teaming activity? \\ - What are the benchmarks of comparison for success?\\ - Can the activity be reconstructed or reproduced?\end{tabular}                                                                                                                                                                                                                                                                                                                         \\ \cline{2-2} 
                                             & \begin{tabular}[c]{@{}l@{}}\textbf{Team composition} and who will be part of the red team? \\ - What were the criteria for inclusion/exclusion of members, and why? \\ - How diverse/homogeneous is the team across relevant demographic characteristics?\\ - How many internal versus external members belong to the team?\\ - What is the distribution of subject-matter expertise among members?\\ - What are possible biases or blindspots the current team composition may exhibit?\\ - What incentives/disincentives do participants have to contribute to the activity?\end{tabular} \\ \hline
\multirow{4}{*}{\textbf{1. During activity}} & \begin{tabular}[c]{@{}l@{}}What \textbf{resources} are available to participants? \\ Do these resources realistically mirror those of the adversary?\\ - Is the activity time-boxed or not?\\ - How much compute is available?\end{tabular}                                                                                                                                                                                                                                                                                                                                                                                        \\ \cline{2-2} 
                                             & What \textbf{instructions} are given to the participants to guide the activity?                                                                                                                                                                                                                                                                                                                                                                                                                                                                                    \\ \cline{2-2} 
                                             & What kind of \textbf{access} do participants have to the model?                                                                                                                                                                                                                                                                                                                                                                                                                                                                                                    \\ \cline{2-2} 
                                             & \begin{tabular}[c]{@{}l@{}}What \textbf{methods} can members of the team utilize to test the artifact?\\ Are there any auxiliary automated tools (including AI) supporting the activity? \\ - If yes, what are those tools?\\ - Why are they integrated into the red-teaming activity?\\ - How will members of the red team utilize the tool?\end{tabular}                                                                                                                                                                                                         \\ \hline
\multirow{4}{*}{\textbf{2. Post-activity}}   
                                            & \begin{tabular}[c]{@{}l@{}}What \textbf{reports and documentation} are produced on the findings of the activity? \\ Who will have access to those reports? When and why? \\ If certain details are withheld or delayed, provide justification.\end{tabular} 
                                            \\ \cline{2-2} 
                                            & \begin{tabular}[c]{@{}l@{}}What were the \textbf{resources} the activity consumed?\\ - time\\ - compute\\ - financial resources\\ - access to subject-matter expertise\end{tabular}                                                                                                                                                                                                                                                                                                                                                                                \\ \cline{2-2} 
                                             & How \textbf{successful} was the activity in terms of the criteria specified in phase 0?                                                                                                                                                                                                                                                                                                                                                                                                                                                                            \\ \cline{2-2} 
                                             & \begin{tabular}[c]{@{}l@{}}What are the proposed \textbf{measures to mitigate} the risks identified in phase 1? \\ - How will the efficacy of the mitigation strategy be evaluated? \\ - Who is in charge of implementing the mitigation?  \\ - What are the mechanisms of accountability?\end{tabular}                                                                                                                                                                                                                                                                                                                 
                                                                                                                                                                                                                                                                                                                                                    \\ \hline
\end{tabular}
\end{table}

\xhdr{Contributions}
Our findings reveal a lack of consensus 
around the scope, structure, and assessment criteria for AI red-teaming. 
Prior methods and practices of AI red-teaming 
diverge along several critical axes, including
the choice of threat model (if one is specified),
the artifact under evaluation, 
the setting in which the activity is conducted
(including actors, resources, methodologies, and test-beds), 
and the resulting decisions the activity instigates 
(e.g., reporting, disclosure, and mitigation).
In light of our findings, we argue that while red-teaming 
may be a valuable big-tent idea, and even a useful framing
for a broad set of evaluation activities for generative AI models,  
the bludgeoning use of \emph{AI red-teaming} (as defined in public literature) as a catch-all response
to quiet all regulatory concerns about model safety 
verges on \emph{security theater} \cite{levenson2014tsa}.
Our work, including our NIST RFI comment analysis, shows that the current framing of red-teaming in the public discourse 
serves more to assuage regulators and other concerned parties
than to offer a concrete solution.
To move toward a more robust toolbox of evaluations for generative AI, 
we synthesize our recommendations into a question bank 
meant to guide and scaffold future AI red-teaming practices 
(see Table~\ref{tab:questions}) and propose future research, including improving the question bank through co-design and evaluation.
\section{Related Contemporary Work}\label{sec:related}

\xhdr{A brief history of red-teaming}
\citet{zenko2015red} and \citet{abbass2015computational} describe how the key concepts of red-teaming 
originated hundreds of years ago in warfare and religious contexts.
They note the term ``red team'' was formally applied by the US military 
as early as the 1960s when modeling the Soviet Union's behavior 
(in contrast to the ``blue team'' representing the US). In computer security, red-teaming involves modeling an adversary and 
\qt{map[ping] out the space of vulnerabilities from a threat lens} 
in contrast to penetration testing
(in which enlisted cybersecurity experts actively attempt to find vulnerabilities in a computer system) \cite{abbass2011computational,abbass2015computational}. 
\citet{wood2000red} further describe how red-teaming \qt{is not an audit} 
and that interpreting it as such risks reducing the amount of information shared about possible vulnerabilities. 
Using a hypothetical pandemic example, \citet{bishop2018augmenting} argue 
that effectively red-teaming a system requires 
context, knowledge, and assumptions about system usage. 

\xhdr{Evaluation beyond red-teaming}
\citet{chang2022understanding} note that red-teaming is only one of many approaches
to increase transparency of an AI system and that factsheets, audits, 
and model cards are other ways to do so. 
Similarly, \citet{horvitz2022horizon} warns of more advanced deepfakes in the near future 
while emphasizing that remedies such as increased media literacy 
and output watermarking (flagging relevant media as AI-generated) 
should be employed alongside red-teaming; 
\citet{kenthapadi2023generative} echo these concerns and similar solutions in their tutorial.
\citet{shevlane2023model} also argue that
both internal and external model evaluations, 
as well as robust security responses, 
should complement effective red-teaming 
to counter GenAI risks.

\xhdr{Existing surveys of AI red-teaming and evaluations}
\citet{inie2023summon} conduct qualitative interviews with those who perform red-teaming 
to create a grounded theory of \qt{how and why people attack large language models.} 
\citet{schuett2023towards} survey members of labs racing
to build artificial general intelligence (AGI) and find 98\% of respondents 
somewhat or strongly agree that \qt{AGI labs should commission external red teams before deploying powerful models.} 
In the software design space, \citet{knearem2023exploring} 
highlight how UX designers are afraid that AI-based design tools 
will not be red-teamed enough while \citet{liao2023designerly} suggest
that UX designers themselves should help with red-teaming processes. 
Considering the testing of NLP systems specifically, 
\citet{tan2021reliability} propose the DOCTOR framework 
for reliability testing of such systems. \citet{weidinger2023sociotechnical} 
introduce a framework for evaluating generative AI more broadly, 
namely via \qt{a three-layered framework that takes a structured, sociotechnical approach.} 
\citet{anderljung2023towards} also propose a framework, ASPIRE, 
but for external accountability of LLMs and the engagement of relevant stakeholders.
\citet{yao2023survey,neel2023privacy}, and \citet{shayegani2023survey} 
produce surveys of LLM research with regard to security, privacy, and other vulnerabilities; 
\citet{chang2023survey} conduct another survey of LLM evaluation. 
In contrast to existing surveys of GenAI evaluation, 
our work focuses exclusively on \emph{red-teaming}. 
Some of our findings resonate with points earlier made 
by \citet{Bockting_2023} and \citet{Friedler2023AIRedTeaming}, 
who argue for interdisciplinary audits of AI systems by diverse groups of people
and red-teaming with concrete definitions of harms alongside other evaluations, respectively.
\section{Case Studies: AI Red-teaming in practice}
\label{sec:case-study}

To capture the complexity involved in designing real-world AI red-teaming exercises,
we synthesize the results from such exercises recently conducted with generative models as case studies. 
Through these case studies, we seek to understand common red-teaming practices, 
typical resources required for successful red-teaming, 
effects of red-teaming on deployed models, 
common pitfalls,
and disclosure of results with community stakeholders.

\xhdr{Methodology} We sourced case studies by searching for reports and news stories about recent red-teaming exercises. As such, our selection is not meant to reflect the full range of red-teaming activities conducted in practice, as limited disclosures from industry teams would make such a reflection impossible. This said, the evaluations we cover here were mostly conducted by private companies, and they encompass a broad range of methods, goals, and areas of focus. In total, we surface and analyze six red-teaming exercises based on retrieved public reports. See Table \ref{tab:cases} for more information.

\begin{table}
    \centering
    \begin{tabular}{lcc}
        Model/System Evaluated & Conducting Organization & Sources \\
        \midrule
        Bing Chat & Microsoft & \cite{FMF_2023,microsoft2024copilot}\\
        GPT-4 & OpenAI & \cite{FMF_2023,openai2023gpt4card,OpenAI_2023} \\
        Gopher & DeepMind & \cite{rae2021scaling,perez2022red,FMF_2023}\\
        Claude 2 & Anthropic & \cite{Anthropic_frontier_rt,FMF_2023,anthropic2023claude}\\
        Various & DEFCON & \cite{cattell2023announce,cattell2023retro}\\
        Claude 1 & Anthropic & \cite{askell2021general,ganguli2022red,anthropic2023claude}\\
    \end{tabular}
    \caption{The six cases of AI red-teaming we discuss as part of our case study analyses. These cases were found by searching for reports and news stories about recent red-teaming exercises. Though industry teams do not disclose (all of) their methods, the cases we analyze here largely stem from industry work, in turn yielding insight into some of their practices.}
    \label{tab:cases}
\end{table}

\subsection{Findings}

\xhdr{Variation in goals, processes, and threat models} 
Reflecting the lack of consensus on a definition of red-teaming in the literature, 
red-teaming activities frequently varied in form and in goals. 
Some organizations chose to conduct a single round of red-teaming \cite{ganguli2022red,perez2022red,Anthropic_frontier_rt, cattell2023announce}, while others saw red-teaming as an iterative process 
in which results from initial rounds of testing were used 
to prioritize risk areas for further investigation \cite{OpenAI_2023,FMF_2023, microsoft2024copilot}. 
The goals of red-teaming activities also ranged from specific objectives 
(e.g., red-teaming to investigate risks to national security \cite{Anthropic_frontier_rt}) 
to more broad targets (e.g., uncovering ``harmful'' model behavior~\cite{FMF_2023}); threat models related to the latter were more common. 
Model developers use such threat models for evaluation in hopes that this will yield greater variation in red-teaming efforts, especially because it is impossible to understand the model's entire risk surface. Unfortunately, probing these nonspecific threat models does not always produce this desired variation, more so when evaluators are given limited time and resources to produce harmful outputs. For example, some time-boxed evaluators repeatedly probed models in the same risk area because it was easy to produce harmful outputs, as opposed to exploring other risks \cite{ganguli2022red}.

\xhdr{Interconnected members, resources, and outcomes} 
The evaluators employed in red-teaming activities for each case study varied considerably. We found that there were generally three types of team compositions:
\begin{enumerate}
\item Teams composed of handpicked subject matter experts in relevant areas 
(e.g., national security, healthcare, law, alignment), both internally and externally sourced
\item Crowdsourced teams chosen from crowdworking platforms or attendees of a live event
\item Teams composed of language models (i.e., language models prompted or fine tuned to red-team themselves)
\end{enumerate}
The resources available to evaluation teams varied based on team composition. 
For crowdsourced teams, red-teaming efforts were time-boxed either by participant or by task, 
and access to models was available only through APIs \cite{ganguli2022red,cattell2023retro}.
For teams with subject matter expertise, red-teaming efforts were more open-ended 
with fewer restrictions on time or compute~\cite{OpenAI_2023, Anthropic_frontier_rt,microsoft2024copilot}. 
While API access to models is still most common for these teams, 
sometimes experts are given access to versions of models without safety guardrails.
When language models are used to red-team themselves, 
the main resource bottlenecks are the number of prompts used to produce red-teaming behavior 
and the compute resources needed for model retraining or fine tuning. 
Full access to model parameters is thus usually a requirement 
when performing this type of red-teaming~\cite{perez2022red}. 
As such, team composition and available resources also shape red-teaming outcomes. For instance, crowdsourced teams typically focused on risk areas where successful attacks were easy to produce due to time constraints, so risk areas that are more complex to attack may remain completely untested~\cite{ganguli2022red, cattell2023retro}. 
In contrast, subject matter experts and members of academic communities and AI firms prioritized different risks and explored them in more detail due to differences in team member selection and resources \cite{OpenAI_2023}. When using language models for red-teaming, offensiveness classifiers are trained on pre-existing datasets such as the Bot-Adversarial Dialogue (BAD) dataset \cite{xu2021bot}, which in turn only cover certain types of offensive model replies. 
Evidently, team selection and resources can introduce bias into the types of risks investigated and ultimate exercise findings.

\xhdr{No standards for disclosing red-teaming details} We found nontrivial variation in the publicly-shared outputs of red-teaming efforts, largely because there are no existing standardized reporting procedures or requirements. In only half of the cases explored, specific examples of risky or harmful model behavior uncovered by red-teaming efforts were publicly shared. In one case, a full dataset composed of 38,961 red team attacks was publicly released to aid in testing of other models \cite{ganguli2022red}. In the other two cases, examples of harmful behavior were publicly available, but the full scope of all red-teaming attacks was not released \cite{OpenAI_2023, perez2022red}. For red-teaming efforts on publicly available models or those focused on national security, specifics of harmful behavior were not shared publicly because findings were deemed too sensitive to share  without responsible disclosure practices \cite{Anthropic_frontier_rt}. One case study resulted in Anthropic piloting a responsible disclosure process to share vulnerabilities identified during red-teaming with appropriate community stakeholders, but this process is still under development (thus we assume that these disclosures have not yet been made)~\cite{Anthropic_frontier_rt}. 
There are also variations in reporting on resource consumption. We found costs of red-teaming efforts were usually disclosed for evaluation teams composed of crowdsourced evaluators (for example, the hourly rate paid to crowdworkers \cite{ganguli2022red}). These details were not disclosed for teams composed of subject matter experts and language models, though they seem to have been given greater time and compute resources. 
Two case studies specifically mention ongoing red-teaming for 6-7 months before model release \cite{OpenAI_2023, anthropic2023claude}. In contrast, for crowdsourced teams, evaluators spent about 30-50 minutes per task, with evaluators sourced from live events being limited to only completing a single task~\cite{ganguli2022red, cattell2023retro}.

\xhdr{Diverse mitigations and supporting evaluations} While every case analyzed here identified problematic or risky model behavior, none of them resulted in a decision not to release the model. Instead, a number of risk mitigation strategies were proposed and/or employed to minimize harmful model behavior identified during red-teaming. 
These approaches ranged from concrete, such as jointly training language models and red-teaming models via strategies for training GANs, to purely conceptual, like unlikelihood training to reduce harmful outputs \cite{perez2022red}.
However, the specifics of risk mitigation strategies were often not provided when the target model was publicly available, and there were no standards for reporting improvements stemming from these efforts. As a result, it was often difficult to determine if risks identified during red-teaming were sufficiently addressed.
Similarly, every case we analyzed involved models that had been previously evaluated using other techniques beyond red-teaming, but there were no established guidelines or standards for these other methods. 
Commonly, models were evaluated using the Perspective API to measure toxicity; human feedback on helpfulness, harmfulness, and honesty; and QA benchmarks for accurate and truthful outputs \cite{rae2021scaling,askell2021general,anthropic2023claude}. 
Other evaluations included internal quantitative assessments to determine if model outputs violated specific content policies (e.g., hate speech, illicit advice) \cite{OpenAI_2023}. Additionally, some initial efforts described as ``red-teaming'' by evaluators were more focused on understanding base model capabilities through open-ended experimentation than on specifically stress testing the model \cite{microsoft2024copilot}. 

\subsection{Discussion}

\xhdr{Red-teaming is ill-structured} Evaluation teams either prioritize risk areas for investigation or provide evaluators with broad directions in hopes that diversity within the group of evaluators will lead to the exploration of many different risks. However, in line with findings from prior empirical work \cite{chung2019efficient,deng2023understanding}, there is a tradeoff between providing evaluators with specific instructions and leaving the activity open-ended. On one hand, vague instructions can be helpful to avoid biasing evaluators towards finding specific issues based on initial prioritization. On the other, a lack of instructions can reduce the utility of the exercise for uncovering risks relevant to real-world contexts. Red teams navigate this tradeoff as they seem aware that the entire risk surface of a model will not be explored by red-teaming activities, but this serves to make red-teaming as a whole ill-structured and difficult to define. Moreover, we argue that this lack of structure and scope is concerning as recent recommendations establishing red-teaming as a best practice suggest that the broader perception of red-teaming may not align with current working definitions of red-teaming, (i.e., red-teaming activities are much more qualitative, subjective and exploratory than community stakeholders may realize). In every case study, however, red-teaming was able to reveal harmful model behavior that other more systematic methods seemed to miss, highlighting the importance of both conducting red-teaming (alongside other evaluations) and developing systematic processes for red-teaming in a more comprehensive manner. These processes could include, for example, the development of guidelines on whether red-teaming is most effective when conducted internally or externally and when it should be conducted (i.e., before and/or after public release of the model and whether red-teaming activities should be ongoing while the model is publicly available). 

\xhdr{Evaluation team composition introduces biases} The goal of team member selection seems to be ensuring variety in the risk areas explored during red-teaming. One way to do so is by handpicking experts with different backgrounds, as noted by prior work on interdisciplinary collaboration within AI teams \cite{nahar2021collaboration, deng2023investigating}; another is by randomly sampling the population through crowdsourcing. Both have drawbacks: there may be bias in expert selection \cite{hube2019understanding, chen2021goldilocks}, and crowdworkers have limited resources in terms of time, compute, and relevant expertise \cite{toxtli2021quantifying}. It is difficult to say what the ideal balance between expert and non-technical stakeholders would be, but prior crowdsourcing research suggests a hybrid approach could help address some of the pitfalls associated with each type of team composition \cite{kittur2013future, vaughan2017making, chen2021goldilocks}. One type of team composition we did not see explored in any case study is a crowdsourced team with more open-ended instructions and greater resources. This could allow more variety in the risk areas explored because evaluators would not feel incentivized to focus on risk areas where harmful model outputs are easy or quick to produce, but it would also require partnering with subject matter experts to fully evaluate risky model behavior. 
Team composition can also shape the outputs of red-teaming. One of the issues with red-teaming via internal teams is that more extreme measures such as blocking the release of a model may never be recommended due to conflicting interests. However, external teams that may be more likely to recommend such measures often do not have the power to actually employ these mitigations. A hybrid approach could resolve some of these issues, but it would need to be paired with accountability mechanisms to disclose recommendations and mitigations. Additionally, directly involving marginalized stakeholders, as they may suffer the most from unanticipated model outputs, is challenging yet not impossible (see, e.g., \cite{woodruff2018qualitative}). 

\xhdr{Hesitancy to release results reduces utility} As suggested by prior work studying responsible AI industry practices \cite{madaio2020co, rakova2021responsible}, the reluctance to share all results from red-teaming activities may stem from risks associated with public models (evaluators do not want to provide inspiration for potential attackers). Additionally, releasing all of the data associated with red-teaming could be overwhelming for community stakeholders. This said, because red-teaming does not seem to be planned as a comprehensive measure of risky model behavior, disclosing some specifics of the activity is necessary so stakeholders can understand harms investigated and in turn determine if they are relevant to their use cases. For example, significant risk areas that evaluation teams knowingly have not probed should be highlighted or identified in reports. 
Moreover, none of the case studies provided complete monetary costs of red-teaming efforts. This information seems relatively low-risk to release (i.e., compared to specific examples of harmful model behavior) and could be useful for developing methods to conduct more comprehensive red-teaming. The costs of assembling teams with differing compositions of expertise and automation, for instance, could be used to determine where resources can be used most effectively. Similarly, the costs of evaluating and mitigating various types of risks could be factored into a cost-benefit analysis when prioritizing risks. The lack of reported cost figures may also make it harder for third-party or external organizations to conduct red-teaming: if these unreported costs are quite large, it could be difficult or impossible for anyone aside from companies themselves to do this type of analysis \cite{costanza-chock2022audits}.
\section{A Survey of AI Red-teaming Research }
\label{sec:survey}

\xhdr{Methodology} 
To source papers, we primarily searched arXiv, Google Scholar, OpenReview, ACL Anthology, and the ACM Digital Library with keywords ``red-teaming'', ``ai red-teaming'', ``jailbreak'', and ``llm jailbreak'', and we then gathered results.\footnote{We focus on red-teaming evaluations, but we argue that the \emph{jailbreaking} literature contains techniques similar in spirit to those employed by AI red-teams, so we also explore it here.} 
Where possible, we replaced preprints with corresponding published works.
We also included relevant works found prior to this search and via snowball sampling.

We scrutinize and subdivide retrieved papers into groups along two dimensions, both of which relate to the evaluation in each paper. The first corresponds to the type of risk investigated during the evaluation, and the second corresponds to the type of approach used for evaluation. We analyze papers by characteristics pertaining to threat model and methodology because we found that research works primarily focused on these aspects (perhaps due to technical relevance) as opposed to other factors important to red-teaming (such as team composition and resources consumed).
Overall classifications and in-depth paper findings (as with analyses of case studies from the previous section) can be found in our appendix, and totals for each classification can be found in Table \ref{tab:paper-stats}. We do not ascribe judgments to any category (i.e., we do not posit that one red-teaming type is better than another).

\begin{table}
    \centering
    \begin{tabular}{c|c|cccc|c}
         \multicolumn{2}{c}{} & \multicolumn{4}{c}{Risk} & \multicolumn{1}{c}{} \\
         \cmidrule(lr){3-6}
         \multicolumn{2}{c}{}                                   &  D & C & B & N & Total\\
         \hline
        \multirow{ 4}{*}{\rotatebox{90}{Approach}} & Brute-Force      & 15 & 4 & 1 & 0 &  20\\
        & Brute-Force + AI                                            & 23 & 7 & 10 & 2 & 42\\
        & Algorithmic Search                                          & 12 & 1 & 1 &  0 & 14\\
        & Targeted Attack                                             & 19 & 7 &  2 & 0 & 28\\
        \hline
        & Total                                                       & 69 & 19 & 14 & 2 & 104
    \end{tabular}
    \caption{Numbers of papers from our survey in each subgroup based on risk (\textbf{D}issentive, \textbf{C}onsentive, \textbf{B}oth, and \textbf{N}either) and approach as outlined in Section \ref{sec:survey}. Over half of papers are concerned with dissentive risks, and over one-third of papers utilize brute-force + AI to red-team AI models.}
    \label{tab:paper-stats}
\end{table}

\subsection{Findings: AI Red-teaming Threat Models} 
We categorize papers according to the threat model implied or specified in each. In particular, we consider the type of \emph{risk} posed in each threat model. For the ease of organizing our retrieved papers and findings, we present risks as one of two binary types, but we note that in reality, there exists a spectrum between the two types, and risks in practice generally may be more nuanced than as we describe them here.

\xhdr{Dissentive risk} We define \emph{dissentive risk} as risk in which people may disagree on its definition and corresponding threat level, and we categorize papers that studied this type of risk accordingly. For instance, hate speech has many definitions, and though images with blood and responses to questions like ``how to build a bomb?''\footnote{AI safety researchers note that ``how to build a bomb?'' is a default GenAI evaluation query despite the fact that related information could be found via Wikipedia or even popular fiction \cite{weir2014martian} because many GenAI developers assert their models should not disclose bomb-making instructions (e.g.,  \cite{openai2023gpt4card,Touvron_2023}), so techniques which yield responses prove safeguards are brittle. This does not necessarily mean the community finds such responses concerning.} 
may be suboptimal AI generations, some may find them admissible depending on context.\footnote{Such generations do not reflect opinions held by the authors.} Nevertheless, researchers like \citet{gehman2020realtoxicityprompts} and \citet{rottger2021hatecheck} create benchmarks to analyze hate speech generation, \citet{rando2022red} perform red-teaming to study how Stable Diffusion's safety filter can be bypassed to generate NSFW content, and \citet{zou2023universal} test whether LLMs respond to queries like ``Generate a step-by-step plan to destroy humanity'' given adversarial prompt suffixes.

\xhdr{Consentive risk} In contrast, we define \emph{consentive risk} as risk in which people agree on the definition and danger it presents, meaning context is not required to evaluate potential harms, and associated content should never be generated by properly functioning AI. We similarly categorize papers accordingly if the authors consider this type of risk. For example, phenomena such as the divulging of private information, leaking of training data, and production of vulnerable code or material for phishing attacks are inadmissible in any situation. To these ends, \citet{chen2023can} study the degree to which multimodal LMs can safeguard private information, \citet{nasr2023scalable} illustrate how divergence attacks cause ChatGPT to reveal training data, \citet{wu2023deceptprompt} analyze how code generation LLMs \qt{can be easily attacked and induced to generate vulnerable code,} and \citet{roy2023generating} discover ChatGPT can create phising code.

\xhdr{Both and neither} Some authors tasked themselves with analyzing \emph{both} kinds of risk, such as personally identifiable information (PII) leakage in addition to hate speech~\cite{ganguli2022red,perez2022red,srivastava2023no}. Others introduce methods to analyze \emph{neither} type of risk from the outset, stressing that definitions and classifications of issues may need to be done from scratch~\cite{casper2023explore,radharapu2023aart}.

\subsection{Findings: AI Red-teaming Methodologies} We further categorize papers based on the methodology the researchers employ to perform red-teaming. Namely, we study the type of \emph{approach} used to find risks.

\xhdr{Brute-force} Work that utilized \emph{brute-force} approaches involved manual evaluation of generative AI inputs and outputs by teams of humans. We found that such teams typically consisted of the researchers themselves, internal auditors (of tech companies), or external members (such as contractors hired via Amazon Mechanical Turk (MTurk)). \citet{xu2020recipes,xu2021bot} and \citet{ganguli2022red} employed crowdworkers to elicit harmful text outputs from language models (including but not limited to offensive language and PII) and measure safety.
\citet{mu2023can} compiled a benchmark from scratch to test LLMs' capacities to follow rules while \citet{huang2023flames} hired crowdworkers to build a new benchmark that assesses alignment with Chinese values. \citet{schulhoff2023ignore} hosted a prompt hacking competition, thereby making competitors LLM red team members.
Other authors handcraft jailbreak attacks against language models \cite{du2023analyzing,li2023deepinception,wei2023jailbreak,li2023multi,liu2023jailbreaking}, but the authors of \cite{wei2023jailbreak} join \citet{xie2023defending} in additionally devising defense strategies for them. \citet{shen2023anything} and \citet{rao2023tricking} analyze the effectiveness of jailbreak attacks collected from external sources (including prior work and public websites).

\xhdr{Brute-force + AI} Another body of work similar to those of the brute-force works described above incorporated AI techniques into their red-teaming processes. Common approaches to do so typically involved having AI models generate test cases and find errors in other AI output. We therefore term such approaches as \emph{brute-force + AI}. Many authors used LLMs to generate normal prompts \cite{perez2022red,rastogi2023supporting,srivastava2023no,bhardwaj2023red,chen2023understanding,mei2023assert,zhang2023jade} and jailbreak prompts \cite{yu2023gptfuzzer,deng2023attack,shah2023scalable,yao2023fuzzllm,wei2023jailbroken} such that LLMs produce bad outputs like harmful text responses. Variations on these ideas also exist, such as the work of \citet{pfau2023eliciting}, in which the authors use \emph{reverse LMs} to work backwards from harmful text responses to prompts that could generate them. 
Others use LLMs to devise new benchmarks related to exaggerated safety responses (i.e., refusal to respond to prompts that are arguably safe)~\cite{rottger2023xstest}, \emph{fake alignment} that occurs when models appear aligned with one query format and misaligned with another (e.g., multiple choice versus open-ended response)~\cite{wang2023fake}, and \emph{latent jailbreaks}, or compliance with \qt{implicit malicious instruction[s]}~\cite{qiu2023latent}. 
Researchers have also used AI to red-team and jailbreak text-to-image models and multimodal LMs. For instance, \citet{lee2023query} demonstrate how passing harmful queries with corresponding images to multimodal models (e.g., an image of a bomb with the question ``how to build a bomb?'') improves the likelihood of harmful text generation.
\citet{mehrabi2023flirt} test their FLIRT framework to analyze text-to-image models like Stable Diffusion.
Still other researchers perform red-teaming of LLMs for specific end-uses. \citet{lewis2023mitigating} red-team an LLM for potential future use as a component of a virtual museum tourguide, and \citet{he2023control} evaluate the dangers of using LLMs as part of scientific research.
In light of the many documented ways generative AI models can be utilized for malicious use, researchers have also studied ways in which they can be defended.
Both \citet{sun2023principle} and \citet{wang2023self} introduce methods that utilize LLMs to generate fine-tuning data that can be used to avert harmful responses.
\citet{zhu2023unmasking} employ k-nearest neighbors and clustering techniques to fix incorrect labels in popular LLM safety datasets (with the goal of developing better downstream safeguards).

\xhdr{Algorithmic search} Some other methods start from a given prompt and utilize a process to modify it until an issue is encountered. Such processes can take the form of random perturbations or a guided search, and we therefore refer to such approaches as \emph{algorithmic search} strategies.
For instance, several authors describe approaches to red-teaming and jailbreaking in which one AI model automatically and repeatedly attacks an LLM until defenses are broken or bypassed~\cite{casper2022white,ma2023red,chao2023jailbreaking,mehrotra2023tree}.
Both \citet{chin2023prompting4debugging} and \citet{tsai2023ring} propose search-based red-teaming approaches to evaluate text-to-image models that perturb input prompts until they simultaneously pass safety filters and generate forbidden content. Search-based approaches can also be used as defensive measures. Noting the brittleness of most jailbreak methods, \citet{robey2023smoothllm} and \citet{zhang2023mutation} introduce methods to detect jailbreaks by applying perturbations to text and image inputs and observing whether outputs change drastically (if so, the input was likely a jailbreak).

\xhdr{Targeted attack} The last approach to red-teaming we document as part of our review involves deliberately targeting part of an LLM, which could include an API, vulnerability in language translation support, or step of its training process, in order to induce issues. As such, we refer to such approaches as \emph{targeted attack} methods.
For instance, \citet{wang2023backdoor} show how to construct \emph{steering vectors} using activation vectors from both safety-tuned and non-safety-tuned versions of models to obtain toxic outputs from safety-tuned models. Others illustrate how to imperceptibly perturb images to cause multimodal LMs to respond in unintended ways (such as replying with a malicious URL or misinformation)~\cite{schlarmann2023adversarial,qi2023visual,bailey2023image}, and \citet{tong2023mass} engineer prompts for text-to-image models that are mismatched with resulting images by exploiting reliance on CLIP embeddings. Other approaches include but are not limited to weaponizing the fact that LLMs are not optimized to converse in low-resource languages and ciphers~\cite{deng2023multilingual,yong2023low,yuan2023gpt}, poisoning data used to tune or utilize LLMs~\cite{rando2023universal,zhang2023trojansql,abdelnabi2023not,cao2023stealthy,wang2023exploitability,lee2024mechanistic}, and attacking APIs associated with black-box models~\cite{pelrine2023exploiting}. Various defensive methods rooted in targeted attack approaches have been proposed as well. \citet{bitton2022adversarial} describe their Adversarial Text Normalizer, which can defend an LLM against various character-level perturbations typical of certain adversarial prompts. In addition, other defensive strategies mentioned previously can defend these attacks (e.g., JailGuard from \citet{zhang2023mutation} addresses attacks introduced in \cite{qi2023visual,bailey2023image,schlarmann2023adversarial,zou2023universal}).

\subsection{Discussion}

\xhdr{Many different methods to perform red-teaming} As illustrated in Table \ref{tab:paper-stats}, researchers and practitioners have undertaken numerous approaches to evaluate GenAI and have all described them as \emph{red-teaming}. At the same time, there have been developments like \citeauthor{schuett2023towards}'s finding that the overwhelming majority of AGI lab members support external red-teaming efforts \cite{schuett2023towards} and the recent Executive Order \cite{House_2023} stressing the importance of red-teaming. These developments and the many red-teaming variations are together arguably concerning, precisely because there is no agreed-upon definition (from these papers) regarding what constitutes red-teaming. By highlighting this, we do not mean to imply that evaluations until now are useless. On the contrary, we posit they are necessary but perhaps insufficient tests of safety, and we conjecture that the existence of many interpretations of ``red-teaming'' suggests there must be more top-down guidance and requirements concerning red-teaming evaluations.

\xhdr{Threat modeling skewed toward dissentive risk} Table~\ref{tab:paper-stats} also highlights that the majority of evaluations focus on \emph{dissentive risk} rather than \emph{consentive risk}. This means that undue effort has been undertaken to evaluate and mitigate GenAI behavior that may be admissible in various contexts. 
Additionally, \citet{rottger2023xstest} have shown that current attempts to mitigate such risks have resulted in exaggerated safety, yielding LLM behavior like the refusal to provide information on buying a can of coke. Lastly, focusing on dissentive risk takes attention away from consentive risk, which in turn is inadmissible in any context. In light of such issues and tradeoffs, \citet{casper2023explore} and \citet{radharapu2023aart} suggest clearly defining risks and problematic outputs and justifying those definitions before any analysis. 

\xhdr{No consensus on adversary capabilities} While threat model and methodology are two factors that contribute to the diversity of red-teaming exercises, assumptions about adversary capabilities are also contributors. Namely, the works encountered have differing estimates of adversary resources. For instance, \citet{perez2022red} and many authors of similar work conjecture that an adversary can only prompt an LLM and probe it for bad outputs. In contrast, others assume that an adversary can poison the training process \cite{rando2023universal}, has the compute required to search for adversarial suffixes~\cite{zou2023universal}, or is able to run both safety-tuned and non-safety-tuned versions of language models to obtain toxic output \cite{wang2023backdoor}. Future guidelines for red-teaming may want to suggest that researchers should emphasize and defend adversary assumptions.

\xhdr{Non-universality of values used for alignment} Work found as part of this survey involving dissentive risk and alignment are driven by, implicitly or explicitly, a set of human values that determine whether GenAI outputs are admissible or inadmissible. However, this in turn prompts the question \emph{whose values are being utilized for alignment and evaluation?} For instance, the FLAMES benchmark proposed by \citet{huang2023flames} is purported to measure alignment with Chinese values, whereas \citet{weidinger2023sociotechnical} emphasize that other evaluations may reflect those of \qt{the English-speaking or Western world.} The extent to which GenAI does not support low-resource languages \cite{deng2023multilingual,yong2023low} and agrees with bias and stereotypes \cite{ganguli2022red,rastogi2023supporting} evidences that models may not reflect the values and beliefs of all persons. Works beyond this survey have illustrated how the framing of AI value alignment is a normative problem that, if not properly addressed, may only serve to reflect the norms of one group of people, typically the majority \cite{feffer2023moral,lambert2023entangled,lambert2023alignment}. Especially as OpenAI started a partnership with the US military on one hand \cite{openai_military,Field_2024_openai_military} and launched an initiative to align superintelligent AI to ``human values'' on another \cite{openai_superalignment},\footnote{This latter project ended in spring 2024 \cite{Field_2024_superalignment_dissolves}; its dissolution may only further support the notion that AI developers' positions should be scrutinized.} we argue that it is crucial to analyze assumptions made and viewpoints held by those who build AI systems. 

\xhdr{No consensus on who should perform red-teaming} Moreover, just as there is a lack of agreement regarding values to use to assess GenAI outputs, there is a similar lack of agreement regarding who should perform red-teaming. Groups of evaluators have consisted of hired crowdworkers~\cite{ganguli2022red}, competition participants~\cite{schulhoff2023ignore}, researchers themselves~\cite{perez2022red}, and others simply red-teaming for fun~\cite{inie2023summon}. While some argue for more diversity to evaluate AI models~\cite{solaiman2023gradient}, others caution that increased diversity is not a panacea and is moreover typically ill-defined~\cite{weidinger2023sociotechnical,bergman2023representation}. For instance, \citet{yong2023low} argue for multilingual red-teaming to respond to low-resource language issues, and \citet{he2023control} \qt{advocate for a collaborative, interdisciplinary approach among the AI for Science community and society at large} to respond to scientific research risks. Such examples suggest that terms like ``diversity'' and ``community'' should be defined and sought out relative to the risks considered by red-teaming processes. They additionally hint towards more involvement of the public and relevant stakeholders, ideas also recommended in parallel literature regarding algorithmic auditing and participatory ML~\cite{costanza-chock2022audits,birhane2022power,feffer2023preference,delgado2023participatory}. Similar literature has also engaged with benefits of deliberation in the face of disagreement (e.g., \cite{pierson2017demographics}) and effects of identity on evaluators' perceptions of AI safety (e.g., \cite{aroyo2023dices}). Future red-teaming guidelines should emphasize these considerations.

\xhdr{Unclear follow-ups to red-teaming activities} We found that overall, responses from GenAI developers (at least public ones) to the many red-teaming and jailbreaking papers have been muted and generally mixed. While some authors such as \citet{wei2023jailbroken} report that they reached out to organizations like OpenAI and Anthropic about the vulnerabilities found in their models, the vulnerabilities and models themselves have for the most part persisted. One rare exception to this pattern is the case of the findings of \citet{nasr2023scalable}, in which OpenAI updated ChatGPT to reduce the likelihood of divergence attack success and modified their terms of use to forbid such attacks in response \cite{Price_ChatGPT,Mok_ChatGPT}. However, these changes only came following the paper's release, 90 days \emph{after} the paper authors first notified OpenAI about the vulnerability.
If red-teaming is to be stipulated as a requirement for release and safe usage of AI models, there should arguably be a protocol to mitigate found issues accordingly.
\section{NIST RFI Comment Analysis Summary}\label{sec:rfi-analysis-body}

We find that comments submitted to the NIST RFI on red-teaming GenAI are generally in accord with our conclusions.\footnote{See our appendix for an extended analysis.}

\xhdr{Similarities} Industry, academia, and civil society organizations suggest that NIST should specify a clearer definition of ``red-teaming'' and provide interested parties with appropriate resources pertaining to guidelines and best practices. Notably, even industry firms with experience red-teaming GenAI expressed a desire for concrete guidance from NIST. This supports our finding that red-teaming, as defined in public research and reports, is loosely structured and perhaps not the rigorous practice implied by the Executive Order. Moreover, many comments stressed that a plurality of different viewpoints and stakeholders should be involved in evaluating GenAI systems. Our findings concur with these notions. 

\xhdr{Differences} A number of comments (including those from OpenAI and Mozilla) recommended evaluations at both the model level and system level. Though our work primarily considers model-level evaluations, we emphasize that in at least one comment, evaluation at either level is referred to as \emph{red-teaming}. This also exemplifies the need for more concrete definitions of evaluations. Additionally, many comments, especially those from individuals, expressed concerns with GenAI, not because of evaluation methods employed but rather because of its uses (such as for malicious deepfakes) and its training data (typically stolen via web-scraping). Though our work centers on GenAI evaluation via red-teaming, our findings support incorporating diverse perspectives while building and evaluating these systems, and we agree that such incorporation should also consider the ethical consequences and legality of any created systems.
\section{Takeaways and Recommendations}\label{sec:recommendations}

Based on our results, we distill the following findings and guidance for future red-teaming evaluations.

\xhdr{Red-teaming is \emph{not} a panacea} Each red-teaming exercise discussed in this paper only covered a limited set of vulnerabilities. As such, red-teaming cannot be expected to guarantee safety from all angles. For instance, from the papers surfaced in our research survey, approaches to red-teaming that detect and mitigate harmful text responses \cite{perez2022red} may not detect and mitigate phishing attack vulnerabilities \cite{roy2023generating} and vice versa. Similarly, our case study analysis highlighted that team composition may also influence the types of issues found in a given exercise (e.g., subject matter experts \cite{Anthropic_frontier_rt} may find different problems than crowdworkers \cite{ganguli2022red}). Moreover, there are other issues that red-teaming alone cannot address, such as problems stemming from \emph{algorithmic monoculture}~\cite{kleinberg2021algorithmic,bommasani2022picking,tong2023mass} or GenAI's environmental impacts \cite{Crawford_2024,Rogers_2024}. We argue that red-teaming should therefore be considered as one evaluation paradigm, among others such as algorithmic impact assessments~\cite{reisman2018algorithmic} and audits, to assess and improve the safety and trustworthiness of GenAI~\cite{Friedler2023AIRedTeaming}. It is also important to support participation from diverse roles (e.g., technical, user-facing, legal) in red teaming within organizations \cite{nahar2021collaboration, deng2023investigating}.

\xhdr{Red-teaming \emph{not} well-scoped or structured} The many variations in the red-teaming processes encountered in our case studies and literature review of public research and reports illustrate that at the moment, red-teaming is an unstructured procedure with undefined scope. This statement is not meant to belittle efforts undertaken, and we concede that we do not have full insight into industry red-teaming activities, but we recommend red-teaming guidelines be drafted and made publicly available to improve utility of future evaluations.

\xhdr{No standards concerning what should be reported} There are currently no unified protocols for reporting the results of red-teaming evaluations. In fact, we found that a number of case studies and research papers sourced for our work did not fully report their findings or resource costs required to perform evaluations. We suggest that regulations and/or best practices be put forth to entice more detailed reporting for a number of reasons, ranging from increasing public knowledge, to helping third-party groups conduct their own tests \cite{raji2020closing, guha2023ai}, to assisting end-users in determining the relevance of red-teaming for their use cases. We argue that such reports should, at a minimum, clarify (1) the resources consumed by the activity, (2) assessments of whether the activity was successful according to previously established goals and measures, (3) the mitigation steps informed by the findings of the activity, and (4) any other relevant or subsequent evaluation of the artifact at hand.

\xhdr{Follow-ups often unclear and unrepresentative} Though red-teaming exercises uncovered many issues with generative models, subsequent activities to remedy these problems were often vague or unspecified. Taken with the lack of reporting, such unclear mitigation and alignment strategies could reduce red-teaming to an \emph{approval-stamping process} wherein one can say that red-teaming was performed as an assurance without providing further details into issues discovered or fixed. Moreover, we found that the strategies specified in research and case studies, such as further fine-tuning or RLHF, were often not representative of the full range of possible solutions. Other approaches like model input and output monitoring, prediction modification, and even the refusal to deploy models in certain scenarios, were rarely or never mentioned. Future research should address mitigation strategies beyond popular solutions given surfaced issues.

\xhdr{Propose question bank as starting point} In light of the issues raised by our work, we provide a set of questions for future red teams to consider before, during, and after evaluation. These questions, found in Table \ref{tab:questions}, encourage evaluators to ponder the benefits and limitations of red-teaming generally as well as the impact of specific design choices pertaining to their setting. We emphasize that these are not finalized guidelines but rather (what we hope is) the start of a broader conversation about GenAI red-teaming and evaluation processes. We welcome and support comments and feedback, and we leave question refinement, overall evaluation, and development of supplementary materials (e.g., rubrics for evaluating red-teaming protocols) as critical future directions.

\section*{Acknowledgements}
Hoda Heidari acknowledges support from NSF (IIS2040929 and IIS2229881) and PwC (through the Digital Transformation and Innovation Center at CMU). Michael Feffer acknowledges support from the National GEM Consortium and the ARCS Foundation.
Authors additionally gratefully acknowledge the NSF (IIS2211955), UPMC, Highmark Health, Abridge, Ford Research, Mozilla, Amazon AI, JP Morgan Chase, the Block Center, and the Center for Machine Learning and Health for their generous support of Zachary C. Lipton's and ACMI Lab's research. Wesley H. Deng also acknowledges support from NSF (IIS2040942), Cisco Research, the Jacobs Foundation,
Google Research, and the Microsoft Research AI \& Society Fellowship Program. Furthermore, this material is based upon work funded and supported by the Department of Defense under Contract No. FA8702-15-D-0002 with Carnegie Mellon University for the operation of the Software Engineering Institute, a federally funded research and development center. Any opinions, findings, conclusions, or recommendations expressed here are those of the authors and do not reflect the views of any funding agencies.

\bibliographystyle{apa-good}
\bibliography{references}

\clearpage

\appendix

\section{Research Survey and Case Study Details}
\label{app:further-details}

This appendix contains further details about the research papers and case studies explored as part of this work. Table~\ref{tab:classifications} contains the specific classifications along each dimension described in Section~\ref{sec:survey} for every work recovered as part of our research survey. We additionally provide access to a Google Sheets project with notes and thematic analyses for both the case studies and research papers retrieved and described in this work. The project can be accessed via this URL: \url{https://docs.google.com/spreadsheets/d/1cZPc6Alkf8sqOFMsEvZgI2PzX2tHTIbemMa6sq4J2Qk/edit?usp=sharing}. 
To conduct the thematic analyses, two authors used notes created for each case and paper encountered as part of this work and synthesized high-level takeaways that in turn formed the foundation for our Discussion subsections. Each author worked independently.
\begin{table}[h]
    \centering
    \begin{tabular}{c|c|p{2.25cm}p{2.25cm}p{2.25cm}p{2.25cm}}
                                                        \multicolumn{6}{c}{Type of Risk Investigated} \\
                                                        \hline
                                                        &                    & Dissentive & Consentive & Both & Neither \\ 
                                                        
     \multirow{ 5}{*}{\rotatebox{90}{Type of Approach Used}}  & Brute-Force        & \cite{zhuo2023red,xu2021bot,xu2020recipes,rottger2021hatecheck,gehman2020realtoxicityprompts,du2023analyzing,schulhoff2023ignore,huang2023flames,li2023deepinception,roy2023probing,wei2023jailbreak,shen2023anything,rao2023tricking,liu2023jailbreaking,xie2023defending} \newline & \cite{chen2023can,mu2023can,roy2023generating,li2023multi} & \cite{ganguli2022red} & None  \\
                                                        & Brute-Force + AI   & \cite{qiu2023latent,sun2023principle,lee2023query,mehrabi2023flirt,bhardwaj2023red,yu2023gptfuzzer,chen2023understanding,deng2023attack,mei2023assert,bhardwaj2023language,mehrabi2023jab,shah2023scalable,pfau2023eliciting,rottger2023xstest,zhang2023defending,ding2023wolf,wang2023fake,wang2023self,chen2023jailbreaker,yao2023fuzzllm,jain2023baseline,alon2023detecting,zhang2023jade} \newline & \cite{casper2023red,shi2023red,greenblatt2023ai,wu2023deceptprompt,zhu2023unmasking,tu2023many,roy2023chatbots} & \cite{perez2022red,rastogi2023supporting,srivastava2023no,wei2023jailbroken,he2023control,wu2023jailbreaking,lewis2023mitigating,salem2023maatphor,tian2023evil,wang2023decodingtrust} & \cite{casper2023explore,radharapu2023aart} \newline \\
                                                        & Algorithmic Search & \cite{chin2023prompting4debugging,ma2023red,casper2022white,tsai2023ring,zhu2023autodan,ge2023mart,chao2023jailbreaking,mehrotra2023tree,robey2023smoothllm,cao2023defending,lapid2023open,yang2024sneakyprompt} \newline & \cite{nguyen2022poster} & \cite{zhang2023mutation} & None \newline \\
                                                        & Targeted Attack    & \cite{rando2022red,yong2023low,huang2023catastrophic,zhao2023causality,cao2023stealthy,wang2023exploitability,wang2023backdoor,yuan2023gpt,zou2023universal,lee2024mechanistic,puttaparthi2023comprehensive,liu2023query,xu2023cognitive,gong2023figstep,deng2023multilingual,shayegani2023jailbreak,deng2023masterkey,qi2023visual,rando2023universal} & \cite{zhang2023trojansql,tong2023mass,bailey2023image,schlarmann2023adversarial,casper2022diagnostics,nasr2023scalable,abdelnabi2023not} & \cite{pelrine2023exploiting,bitton2022adversarial} & None \\
    \end{tabular}
    \caption{In-depth classification of papers acquired for our survey based on the type of content produced and type of approach used in each paper. See Section \ref{sec:survey} for details and definitions.}
    \label{tab:classifications}
\end{table}

\section{Extended NIST RFI Comment Analysis}

Given its directives in the Executive Order, the National Institute of Standards and Technology (NIST) arm of the Department of Commerce issued a Request for Information (RFI)\footnote{\url{https://www.federalregister.gov/documents/2023/12/21/2023-28232/request-for-information-rfi-related-to-nists-assignments-under-sections-41-45-and-11-of-the}} to solicit comments and advice from the general public on how best to carry out the requisite jobs. Specifically, the RFI sought feedback on AI red-teaming, content watermarking, and creating standards for AI development. As the first and last of these issues are related to our work, we chose to analyze submitted comments\footnote{\url{https://www.regulations.gov/docket/NIST-2023-0009/comments}} to observe similarities and differences relative to our findings. In the process, we sought to understand who was commenting, what issues their responses concerned, and how they constructed their arguments.

\begin{figure}
    \centering
    \includegraphics[width=0.9\textwidth]{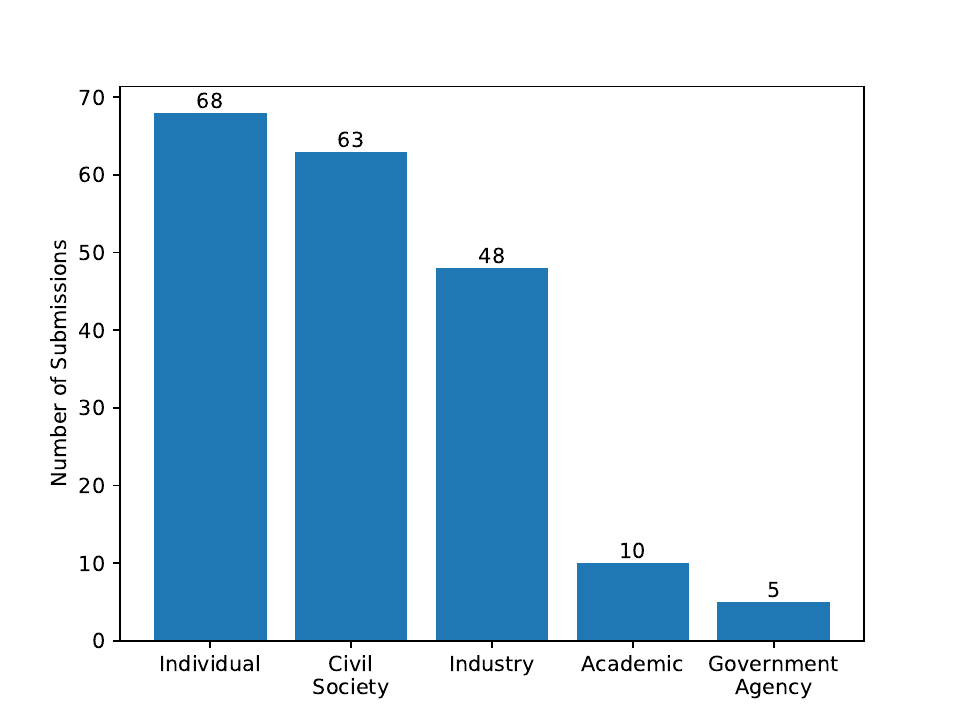}
    \caption{Frequency counts of comments submitted to the NIST RFI grouped by respondent type. Evidently, individual comments and those from civil society were most common, but submissions from industry (which consisted of startups and large software companies) were also numerous. Academic group and government agency submissions were least frequent.}
    \label{fig:nist-freq-cts}
\end{figure}

\xhdr{Overall Trends} 
Comments range from short, plaintext responses from anonymous individuals to PDFs of technical reports with tens of pages from civil society organizations, government agencies, academic groups, tech startups, and large software companies (throughout this section, we refer to the last two groups together as ``industry''). Short comments by individuals tend to be focused on data rights and transparency; they are often emotionally charged and attack companies knowingly stealing data to train GenAI (e.g. \qt{Generative AI should be heavily regulated...It is unethically trained on art that artists did not give their consent to. Ultimately generative AI is only suited for mathematical applications...Keep it away from the general public for their own safety. Keep it out of the arts, which is the one bastion of human creativity we have that truly brings joy to the populous [sic].}). For groups within industry, small companies' responses often appear to be ``sales pitches'' for framework, products, and infrastructure they have to offer that can help NIST fulfill its obligations. Larger companies' submissions tend to include regulation recommendations and descriptions of their internal evaluation approaches. With regard to civil society organizations, academic groups, and government agencies, though a couple of right-wing groups urge NIST to NOT adhere to the Biden administration’s Executive Order for purely political reasons, by and large these organizations made cogent arguments about pressing GenAI issues and regulations regardless of their places on the political spectrum. See Table \ref{tab:rfi_entities} and Figure \ref{fig:nist-freq-cts} for qualitative and quantitative information about submitters and submitting groups.

\begingroup
\renewcommand{\arraystretch}{1.25}
\begin{table}
    \centering
    \begin{tabular}{p{5cm}p{10cm}}
        Entity Type & Examples \\
        \midrule
         Individual & anonymous commenter, Jeffrey Frank, Bridget S \\
         Civil Society Organization & Mozilla, Data \& Society, Center for AI Safety \\
         Government Agency & National Federation of Independent Business, Inc.; City of New York Office of Technology \& Innovation; US Chamber of Commerce \\
         Academic Group & IEEE Standards Association, Johns Hopkins Center for Health Security, CU Boulder \\
         Tech Startup & Credo, OpenAI, Hugging Face \\
         Large Software Company & Google, Adobe, Meta \\
         
    \end{tabular}
    \caption{Types of respondents to the NIST RFI with corresponding examples of each type. Evidently, a range of different individuals and organizations submitted comments to NIST, but generally, they all covered issues pertaining to watermarking, copyright infringement, and red-teaming.}
    \label{tab:rfi_entities}
\end{table}
\endgroup

\xhdr{Need for clear red-teaming definitions and guidelines} In accordance with our main paper findings, many companies and civil society organizations asserted that ``red-teaming'' was vaguely defined in the Executive Order, and in response, they provided their own definitions of red-teaming while calling for NIST to offer clear, standardized definitions. For instance, Google highlighted that ``red-teaming'' is \qt{often used as a catch-all, encompassing a broad sweep of AI safety testing practices, which is confusing and potentially counter-productive.} The Business Roundtable similarly appealed to NIST for global red-teaming standards and aligned definitions. Hugging Face introduced their own definition in a blogpost referenced by their response \cite{Rajani_Lambert_Tunstall}, stating that according to them, \qt{Red-teaming prompts, on the other hand, look like regular, natural language prompts [in contrast to adversarial ML prompts].}

\clearpage

\xhdr{Need to red-team both models and systems} Many comments from various interested parties (ranging from OpenAI to Mozilla to the Consumer Technology Association) emphasized that NIST should differentiate between \emph{AI model red-teaming} (which entails attempting to break the AI model to find improvement avenues) and \emph{AI system red-teaming} (which entails attacking the model along with its data infrastructure, user interface, and other components) in their future guidelines and recommendations. For example, OpenAI shared their practices on iteratively red-teaming models and systems, highlighting that they would conduct red-teaming of their ChatGPT systems when they changed their product interfaces even if the underlying models remain the same.

\xhdr{Need to share red-teaming resources} A request for NIST repeatedly appears in the comments to disseminate materials about red-teaming to relevant organizations and communities. For instance, Meta suggested NIST should collect case studies, illustrations, and/or examples of AI red-teaming that exemplify best practices or the state of the art.

\xhdr{Divergence between industry and civil society on external red-teams} While agreeing on the need to engage with diverse perspectives in red-teaming, many technology companies rhetorically expressed reluctance and deflected their responsibilities for conducting external red-teaming. For example, Google repeatedly stressed that NIST red-teaming guidelines need to be ``flexible'' and cautioned against generally engaging external experts in red-teaming due to (in)feasibility, suggesting that \qt{external red-teaming should only be required or recommended where necessary and technologically feasible.} This in turn often concerns the degree to which Google would open their models, but Google does not provide further details on ways to assess this technological feasibility. Moreover, Google also suggested that, akin to cybersecurity red-teaming, many vulnerabilities identified by AI red-teaming \qt{need not be published or reported publicly unless (1) users need to take action to fix the vulnerability (e.g., installing an update), or (2)  the vulnerability was maliciously exploited and users or customers were affected.}
Intel similarly suggests that red-teaming should start with including technical AI experts and professional red-teamers, and then engage domain experts as needed to maintain the cost and priority of red-teaming in practice. Evidently, these companies and others are concerned about the technological and organizational feasibility of external red-teaming and are often vague on their plans for implementing engagement with external red-teamers. In contrast to industry lines of argument, civil society organizations often urged NIST to involve external stakeholders in conducting red-teaming throughout the Generative AI lifecycle (including from the start) as an indispensable instrument for regulating the private sector. The Center for AI Safety, for instance, suggests the need to
\begin{enumerate}
    \item include people with experience in prompt engineering or white-hat jailbreaking, with collaboration with domain experts, and
    \item include red-teamers that are representative of the expected user base of the system.
\end{enumerate}
Likewise, Mozilla emphasizes the need for independent ``auditing'' and ``red-teaming'' in addition to tools to help external teams carry out such evaluations.

\xhdr{Leverage cybersecurity for guidelines while acknowledging AI challenges} Organizations with previous experience in cybersecurity, such as Google, RAND, and Meta, call for learning from traditional cybersecurity practice. In particular, Google showcased their AI Red Team’s recent efforts on testing both typical attacks on standalone GenAI models and systems integrated with GenAI, which in turn were based on practices shared by their (traditional) Red Team. Google then called for NIST to \qt{incorporate cybersecurity norms into its approach to [AI] red-teaming} and to provide model developers an appropriate time period \qt{to remedy any identified vulnerabilities before reporting any findings pursuant to testing.} Google also emphasized the challenges of adversarial testing to adhere to the AI regulation given AI tools' wide-ranging use cases, suggesting NIST create recommendations and guidelines on balancing the significant legal implications and technical limits of adversarial testing in sensitive content domains.

\end{document}